\documentclass[twocolumn,superscriptaddress,showpacs,preprintnumbers,amsmath,amssymb]{revtex4}

\usepackage{graphicx, color}
\usepackage{dcolumn}
\usepackage{bm}
\usepackage{wasysym}

\begin{document}


\title{A mean field study of single-particle spectra evolution
in $Z=14$ and $N=28$ chains\\}

\author{Dimitar Tarpanov}
 \affiliation{Institut de Physique Nucl\'eaire, IN2P3-CNRS and
 Universit\'e Paris-Sud, F-91406
Orsay Cedex, France}
\affiliation{Institute for Nuclear Research and Nuclear Energy,
1784 Sofia, Bulgaria}

\author{Haozhao Liang}
\affiliation{Institut de Physique Nucl\'eaire, IN2P3-CNRS and
    Universit\'e Paris-Sud, F-91406
    Orsay Cedex, France}
\affiliation{State Key Laboratory of Nuclear Physics {\rm\&} Technology,
    School of Physics, Peking University, Beijing 100871, China}

\author{Nguyen Van Giai}
\affiliation{Institut de Physique Nucl\'eaire, IN2P3-CNRS and
 Universit\'e Paris-Sud, F-91406
Orsay Cedex, France}

\author{Chavdar Stoyanov}
\affiliation{Institute for Nuclear Research and Nuclear Energy,
1784 Sofia, Bulgaria}

\date{\today}

\begin{abstract}
We study the mechanisms which reduce the proton
$1d_{3/2}$-$1d_{5/2}$ spin-orbit splitting and the neutron
$1f_{7/2}$ subshell closure in $^{42}$Si. We use various
self-consistent mean field models: non-relativistic
Skyrme-Hartree-Fock and relativistic density-dependent
Hartree-Fock. Special attention is devoted to the influence of a
tensor component in the effective interaction.
It is found that the tensor force indeed governs the reduction of the 1d proton
spin-orbit splitting. On the other hand, the reduction of the
neutron $1f_{7/2}$ subshell closure is not clearly related to the
tensor force.
\end{abstract}

\pacs{...}
\maketitle

\section{Introduction}
With the progress in experimental studies of single-particle nuclear
spectra it has become clear that the single-particle levels undergo
modifications when the neutron number $N$ or proton number $Z$
change. From the experimental side, these studies are a difficult
task because the nuclei involved are generally short-lived and their
production rate can be low. It is important to have in parallel
theoretical models which may serve as guidelines for exploring the
evolution of nuclear properties with $N$ and $Z$ numbers. A recent
example is the work of Ref. \cite{Otsuka05} drawing the attention on
the possible effects of the neutron-proton tensor force on
single-particle spectra along an isotopic or isotonic chain. This
has triggered a number of experimental \cite{Gaudefroy06,Bastin07}
and theoretical \cite{Colo07,Brink07,Lesinski07,Zou08}
works exploring this issue.

In the present work we would like to use the self-consistent mean
field approach to analyze two findings based on experimental data
complemented by a shell model analysis \cite{Gaudefroy06,Bastin07}:
1) the shrinkage of the proton spin-orbit splitting
$1d_{3/2}$-$1d_{5/2}$ in the $Z=14$ isotopes when going from $N=20$
($^{34}$Si) to $N=28$ ($^{42}$Si); 2) the quenching of the neutron
shell closure gap $2p_{3/2}$-$1f_{7/2}$ when going along the
isotonic chain $N=28$ from $Z=20$ ($^{48}$Ca) to $Z=14$ ($^{42}$Si). Our
analysis is based on both non-relativistic Hartree-Fock-BCS (HF-BCS)
approach with Skyrme-type interactions \cite{Chabanat98b} and
relativistic Hartree-Fock-BCS (RHF-BCS) approach \cite{Long06}. Our
aim is to discuss the effects, among other things, of the tensor
component of the two-body effective interaction. The HF-BCS model is
well suited for that purpose with the several Skyrme-type tensor
forces proposed recently. On the relativistic side, the widely used
relativistic mean field (RMF) model cannot help because it is just a
Hartree approximation where a tensor component in the interaction
would have no effect. Therefore, we use in our analysis the RHF
model where exchange (Fock) terms are kept and the pion-induced
tensor interaction can play its role \cite{Long08}.


This paper is organized as follows. In Sec.II we briefly give the
main ingredients of the two models used in this analysis, namely
the non-relativistic HF-BCS model with a Skyrme type interaction,
and the relativistic RHF-BCS model with density-dependent
effective Lagrangians. In Sec.III the results obtained for the
$Z=14$ isotopic chain and the $N=28$ isotonic chain of nuclei are
discussed. Conclusions are drawn in Sec.IV.

\section{Self-consistent mean field models}
We assume a spherical symmetry description of the nuclei studied
here. Although this is a strong assumption for some nuclei of the
isotopic and isotonic chains that we consider, it allows an easier
insight into the role of separate parts of the mean field such as
the central and spin-orbit potentials.

\subsection{HF-BCS mean field with Skyrme interaction}
The Skyrme-HF model is widely used and we refer to
Ref. \cite{Chabanat98b} for the notations of the present work as
well as for the effective interaction SLy5 used here for
generating the self-consistent mean field. In addition to the
usual central and spin-orbit components of the Skyrme interactions
(represented here by the SLy5 parametrization) we want to study
the possible effects of a tensor component of the Skyrme force.
This component was introduced in earlier versions of the Skyrme
force \cite{Skyrme56,Stancu77,Liu91} but it is only recently that
attention was focused on its effects on single-particle spectra in
spin-unsaturated nuclei.

Along the chain of $N=28$ isotones ($Z=14$ isotopes) the proton-proton
(neutron-neutron) pairing correlations will play a role, and they
are treated in BCS approximation \cite{Ring-Schuck} using a
zero-range, density-dependent pairing interaction of the form:
\begin{eqnarray}
V^{(\rm{n~or~p})} &=& V_{0}^{(\rm{n~or~p})} \left(1 -
\frac{\rho({\boldsymbol R})}{\rho_0}
\right)\delta({\boldsymbol r}_1 - {\boldsymbol r}_2)~, \label{eq1}
\end{eqnarray}
where ${\boldsymbol R}=({\boldsymbol r}_1 + {\boldsymbol r}_2)/2$, $\rho_0=0.16$ fm$^{-3}$ is
the nuclear matter saturation
density and $V_{0}^{(\rm{n~or~p})}$ is fitted for each chain to
reproduce the empirical two-neutron(two-proton) separation
energies. The pairing window contains the orbitals of the $1p$, $2s$-$1d$ and $2p$-$1f$ shells.

The effect of a tensor component in the Skyrme interaction is very
simply described by a modification of the coefficients $\alpha$
and $\beta$ which multiply neutron or proton spin densities as
explained, e.g., in Ref. \cite{Colo07}. This modification affects
the single-particle spin-orbit potentials, especially in the case
of spin-unsaturated subshells which contribute importantly to the
spin densities. One has:
\begin{eqnarray}
\alpha &=& \alpha_C + \alpha_T~,\nonumber \\
\beta &=& \beta_C + \beta_T~, \label{eq2}
\end{eqnarray}
where $\alpha_C, \beta_C$ come from the central,
velocity-dependent part of the Skyrme force whereas $\alpha_T,
\beta_T$ come from the tensor component. Using the Skyrme force
SLy5 the values of $\alpha_T$  and $\beta_T$ were
determined \cite{Colo07} to be $\alpha_T=-170$ MeV fm$^5$,
$\beta_T=100$ MeV fm$^5$, and we shall adopt these values.

In the non relativistic Skyrme-HF model the radial HF equations
can be expressed in terms of an energy-dependent equivalent
potential $V_{\rm{eq}}^{lj}$:
\begin{equation}
\frac{\hbar^2}{2m} \left[-\frac{d^2}{dr^2} \psi(r)
+\frac{l(l+1)}{r^2} \psi(r) \right] + V_{\rm{eq}}^{lj} (r,\epsilon)
\psi(r)=\epsilon \psi(r)~, \label{MG1}
\end{equation}
where
\begin{equation}
V_{\rm{eq}}^{lj}(r,\epsilon) =\frac{m^*(r)}{m} U_0 (r) + \frac{m^*(r)}{m}
U_{\rm{so}}^{lj}(r) + V_{\rm{eq}}^{\rm{m*}}~.
\label{MG2}
\end{equation}
Here, $U_{\rm{so}}^{lj}(r)=U_{\rm{so}}(r) \times [j(j+1)-l(l+1)-3/4]$,
$U_{\rm{so}}(r)$ is  the spin-orbit HF potential,
\begin{equation}
\begin{split}
V_{\rm{eq}}^{\rm{m*}} = \left[1-\frac{m^*(r)}{m} \right] \epsilon
-\frac{m^{*2}(r)}{2m\hbar^2} \left( \frac{\hbar^2}{2m^*(r)}
\right)^{'2}  \\ + \frac{m^*(r)}{2m} \left(
\frac{\hbar^2}{2m^*(r)} \right)^{''} ~,
\end{split}
\label{MG3}
\end{equation}
and $U_0(r)$ and $m^*(r)$ are the HF central potential and
effective mass, respectively \cite{Chabanat98b}. The Coulomb
potential is included in $U_0$, in the case of protons.
With the help of Eqs. (\ref{MG1})-
(\ref{MG3}) we can write $\epsilon$ as a sum of kinetic, central,
spin-orbit and non-locality (or $m^*$-dependent) contributions:
\begin{equation}
\epsilon = \epsilon_{\rm{kin}} + \epsilon_{\rm{cen.}} + \epsilon_{\rm{s.o.}} +
\epsilon_{m*}~. \label{MG4}
\end{equation}
In the above expression the three last terms correspond to the
three terms of Eq.(\ref{MG2}). This decomposition is useful for
understanding the evolution of single-particle energies when $N$ or
$Z$ is changing \cite{Grasso07}, and it will be used in the
discussion of results in Sec.III.

\subsection{RHF-BCS mean field with density-dependent Lagrangians}
Another type of mean field approach is based on covariant
effective Lagrangians. A very popular and successful version is
the RMF (see, e.g., Ref. \cite{Ring96}). However, the RMF is only a
Hartree approximation and it cannot describe the effects of a
two-body tensor interaction brought about by exchanging pions, or
rho mesons with tensor coupling because these effects would appear
in the exchange (Fock) contributions to the total energy. Thus,
the RMF aproach is not suitable for our present purpose.

Recently, progress have been made with the RHF-BCS approach. New
density-dependent effective Lagrangians have been proposed and
they are able to give a satisfactory overall description of bulk
properties of nuclei throughout the mass
table \cite{Long06,Long06a}. In this work we will compare results
obtained with two different effective Lagrangians: the parameter
set PKO1 \cite{Long06} which includes $\sigma$, $\omega$,
$\rho$-vector and $\pi$ exchanges, and the parameter set
PKO2 \cite{Long06a} which does not include $\pi$ exchange. Thus, the
comparison of the predictions of PKO1 and PKO2 can give an idea of
the effects on single-particle spectra due to a tensor component
in the interaction, since the Fock terms of the one-pion exchange
are indeed sensitive to this tensor component.

The two effective Lagrangians PKO1 and PKO2 are treated in the
RHF-BCS approximation within a spherical symmetry assumption as in
the Skyrme HF-BCS case. The pairing interaction has the same
functional form as in Eq.(\ref{eq1}), and the interaction
parameters $V_{0}^{(\rm{n~or~p})}$ are adjusted as in the
non-relativistic case. The RHF-BCS equations are self-consistently
solved in coordinate space \cite{Long06,Long06a}.

\section{Results and discussion}
\subsection{The $2s$-$1d$ proton spectra in $Z=14$ isotopes}
We first examine the evolution of the proton single-particle
spectra of the $2s$-$1d$ shell in the isotopic chain $Z=14$ when the
neutron number increases from $N=20$ ($^{34}$Si, $\nu 1f_{7/2}$
empty) to $N=28$ ($^{42}$Si, $\nu 1f_{7/2}$ filled). The question
of reduction of the proton $1d_{3/2}$-$1d_{5/2}$ spin-orbit splitting when
going from $^{34}$Si to $^{42}$Si is particularly interesting. In
a shell model analysis of the data it is found \cite{Bastin07} that
this spin-orbit splitting is reduced by about 1.94~MeV if one
fills the $1f_{7/2}$ neutron subshell.

\begin{figure*}[htbp]
\includegraphics[width=0.7\textwidth]{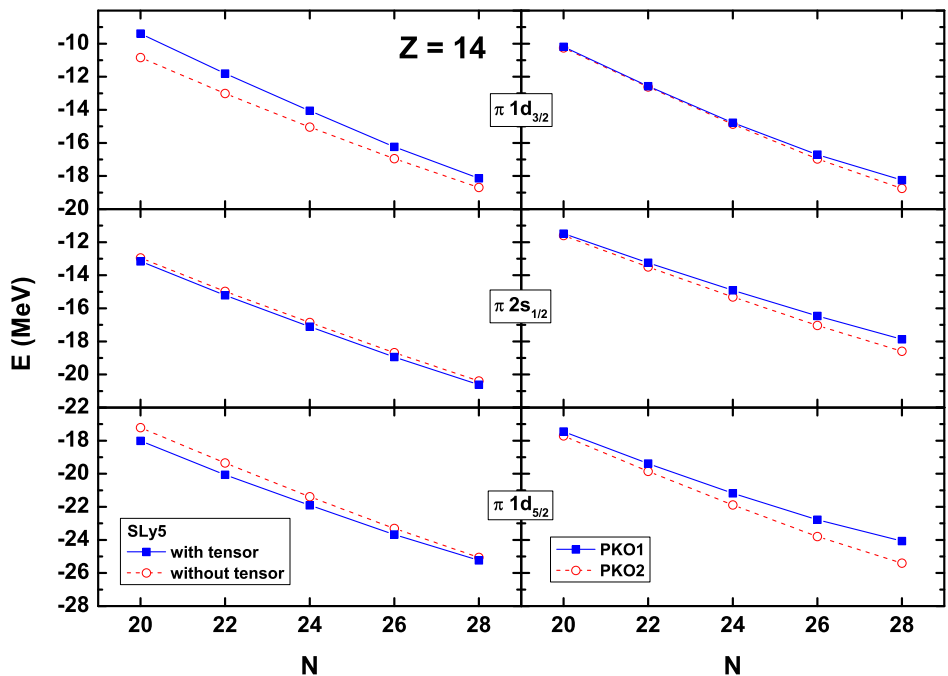}
\caption{The $2s$-$1d$ proton single-particle levels in the Si isotopic
chain calculated in the mean field approach. Left panel:
non-relativistic model with SLy5; right panel: relativistic model
with PKO1 and PKO2.}  \label{Fig1}
\end{figure*}

In the left panel of Fig.~\ref{Fig1} we present the $2s$-$1d$ proton levels
calculated with the Skyrme force SLy5, with and without the tensor
component of Ref. \cite{Colo07}. Corresponding results obtained
with the RHF-BCS model are shown in the right panel, using the
PKO1 (with pion exchange, i.e., with tensor coupling) and PKO2
(without pion exchange) parametrizations. The general down-going
trend of all results as a function of $N$ is easily understood as an
effect of the neutron-proton symmetry potential.

\begin{table}[h]
 \caption{\label{Tab:1} The $1d_{3/2}$-$1d_{5/2}$ splitting for for the nuclei $^{34-42}$Si
    calculated for the models described in the text; The units are MeV.}
 \begin{ruledtabular}
 \begin{tabular}{cccccc}
 ($N$,$Z$) & (20,14) & (22,14) & (24,14) & (26,14) & (28,14)    \\
\hline
SLy5 & 6.38 & 6.33 & 6.35 & 6.35 & 6.36 \\
SLy5+T & 8.62 & 8.26 & 7.85 & 7.43 & 7.10 \\
PKO1 & 7.26 & 6.81 & 6.41 & 6.07 & 5.83 \\
PKO2 & 7.46 & 7.24 & 7.02 & 6.82 & 6.65 \\
 \end{tabular}
 \end{ruledtabular}
 \end{table}

The calculated values of the proton spin-orbit splitting,
$\Delta_{\rm{so}}(N)\equiv \epsilon_{1d_{3/2}}-\epsilon_{1d_{5/2}}\vert_N$
are shown in Table~\ref{Tab:1}.
It can be seen that the reduction of
$\Delta_{\rm{so}}(N)$ from $N=20$ to 28 is practically zero for SLy5 but
it becomes 1.52~MeV for SLy5 plus tensor. For PKO1 and PKO2 the
reduction is 1.43~MeV and 0.81~MeV, respectively. It seems that
the tensor force helps to reduce $\Delta_{\rm{so}}(N)$ in both the
non-relativistic 
and relativistic models.

\begin{figure}[htbp]
\includegraphics[width=0.45\textwidth]{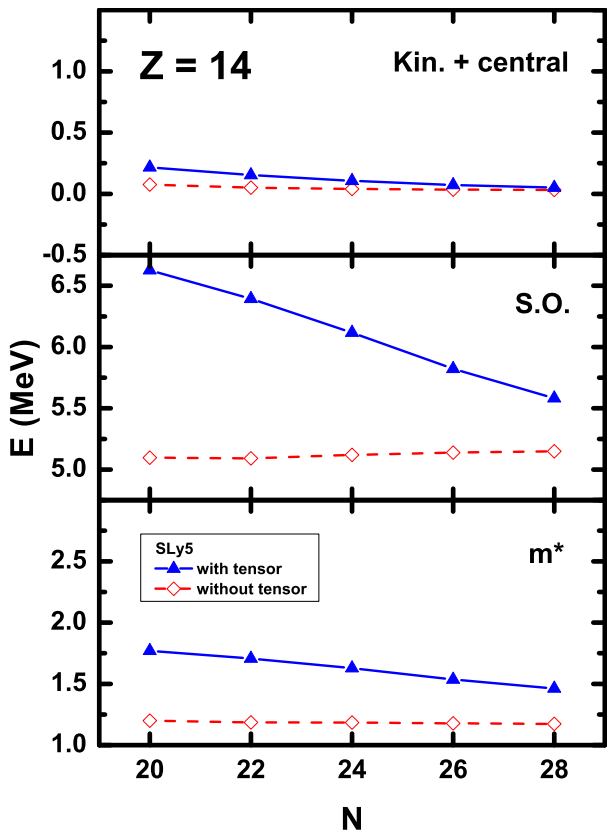}
\caption{The contributions to $\Delta_{\rm{so}}(N)$ according to the
decomposition Eq.(\ref{MG4}). The results correspond to the Skyrme
interaction SLy5, with (solid lines) and without (dashed lines)
tensor component. } \label{Fig2}
\end{figure}

Actually, the two relativistic models PKO1 and PKO2
differ not only by the pion-coupling term but also by the
strengths of other meson-nucleon couplings and therefore, one
cannot relate the changes in $\Delta_{\rm{so}}(N)$ only to the tensor
coupling. On the other hand, it is an easier task to analyze the
evolution of $\Delta_{\rm{so}}(N)$ in the non-relativistic model. We
use the decomposition of Eq.(\ref{MG4}) to cast $\Delta_{\rm{so}}(N)$
into a sum of 3 terms: 1) a kinetic plus central contribution; 2)
a spin-orbit contribution; 3) a term due to $m^*(r)\le m$ and
related to the non-locality of the mean field. These three terms
are shown in Fig.~\ref{Fig2}. The kinetic and central contributions are of
opposite signs and comparable magnitudes, their sum remains small
from $N=20$ to $28$ and the tensor force has little effect on it.
The spin-orbit term is nearly flat for SLy5 whereas the results of
SLy5 + tensor show a strong decrease of about 1 MeV. This effect
is easily understood because of the properties of the Skyrme-HF
mean field and of the neutron-proton tensor interaction. Firstly,
the neutron spin density $J_n$ becomes larger when one fills the
neutron orbital $1f_{7/2}$ \cite{Chabanat98b}. Then, the proton
spin-orbit potential which contains a term $\beta_T J_n$ changes
accordingly. Since the neutron $1f_{7/2}$ orbital is of the type
$j=l+1/2$, the sign of this change is such that the proton $1d_{5/2}$
state is pushed up and $1d_{3/2}$ is pushed down according to the
general rule of matrix elements of a tensor
interaction \cite{Otsuka05}.
The proton spin density remains the same for the entire chain,
so the term  $\alpha_T J_p$ contributes in a same way to all nuclei.
Finally, the non-locality term is flat
for the SLy5 results but it decreases by about 300 keV from $N=20$
to 28 for the SLy5 + tensor case because it is practically
proportional to $\Delta_{\rm{so}}(N)$.

Thus it can be concluded that, in the framework of the
self-consistent mean field with Skyrme interactions, the reduction
of the proton $1d_{3/2}$-$1d_{5/2}$ spin-orbit splitting in $^{42}$Si can
be attributed to the effect of a tensor component in the effective
interaction. For the RHF-BCS model, one observes that the
reduction is also enhanced by a tensor component brought about by
a pion-induced interaction. However, all of the present discussion
is done strictly in the framework of a static mean field approach,
and it must be kept in mind that effects beyond mean field like
 particle-vibration coupling \cite{Mahaux85} have to be evaluated for these
nuclei. Because of these missing effects one cannot compare
directly our calculated spectra with empirical ones. Nevertheless,
the predictions of the mean field approach concerning evolutions
of energy differences remain meaningful.

\subsection{The $2p$-$1f$ neutron spectra in $N=28$ isotones }
We now discuss the changes of neutron single-particle energies in
the $2p$-$1f$ shell when going along the $N=28$ isotones from $^{48}$Ca
($Z=20$) to $^{42}$Si ($Z=14$).

\begin{figure*}[htbp]
\includegraphics[width=0.7\textwidth]{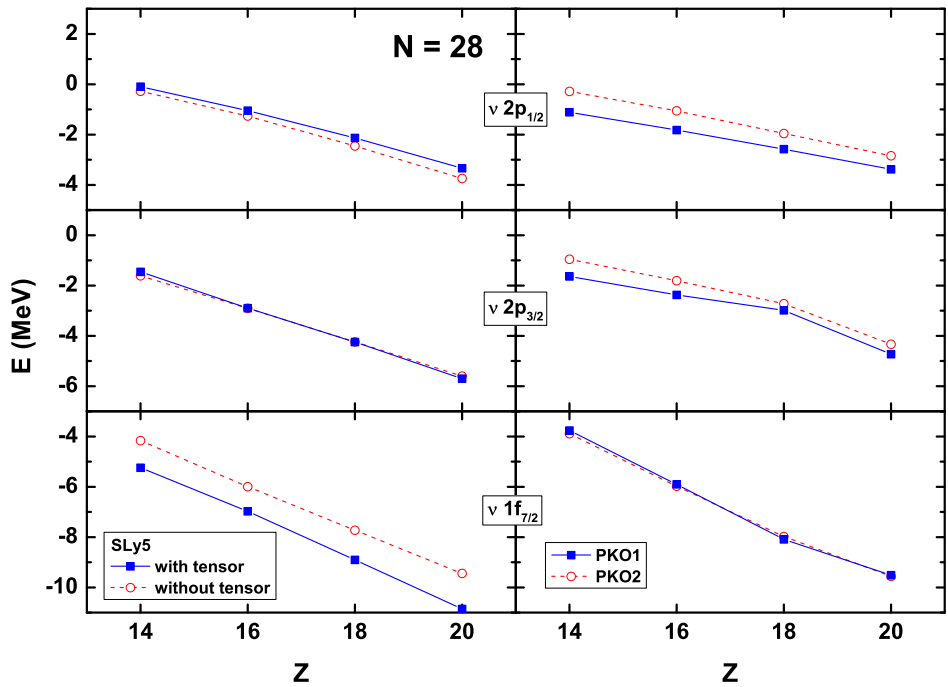}
\caption{The $2p_{1/2}$, $2p_{3/2}$ and $1f_{7/2}$ neutron
single-particle levels in the $N=28$ isotonic chain calculated in the
mean field approach. Left panel: non-relativistic model with SLy5;
right panel: relativistic model with PKO1 and PKO2.}  \label{Fig3}
\end{figure*}
Experimental observations combined with a shell model
analysis \cite{Gaudefroy06,Bastin07} lead to the conclusion that a
compression of the neutron $2p$-$1f$ spectrum occurs when one removes
protons from the $1d_{3/2}$ and $2s_{1/2}$ orbitals, keeping $N=28$
fixed. This compression could affect by about 1 MeV the distance
between the neutron $1f_{7/2}$ and $2p_{3/2}$ states and reduce
the $N=28$ gap from 4.8 MeV in $^{48}$Ca to about 3.8 MeV in
$^{42}$Si.

In Fig.~\ref{Fig3} we present the energies of the three lowest states of
the neutron $2p$-$1f$ shell calculated with the non-relativistic and
relativistic mean field models. We observe again the down-going
trends as functions of $Z$, a manifestation of the increasing
symmetry potential in the mean fields. Looking at the $N=28$ gap
$\Delta_{\rm{gap}}(Z)\equiv \epsilon_{2p_{3/2}}-\epsilon_{1f_{7/2}}\vert_Z$
we can see from Fig.~3 that, in the non-relativistic case adding a
tensor force affects little the $2p_{3/2}$ state but pulls down
the $1f_{7/2}$ state and consequently increases $\Delta_{\rm{gap}}(Z)$
for all $Z$ values. On the other hand, in the relativistic case the
position of the $1f_{7/2}$ state is almost the same for both PKO1
and PKO2 parametrizations but the $2p_{3/2}$ state is lower with
PKO1 (with tensor component) than with PKO2, so that the model
with a tensor component seems to lead to smaller
$\Delta_{\rm{gap}}(Z)$. However, as noticed above, the differences
between PKO1 and PKO2 are not only in the tensor component
contrarily to SLy5 and SLy5 + tensor.
\begin{figure}[htbp]
\includegraphics[width=0.45\textwidth]{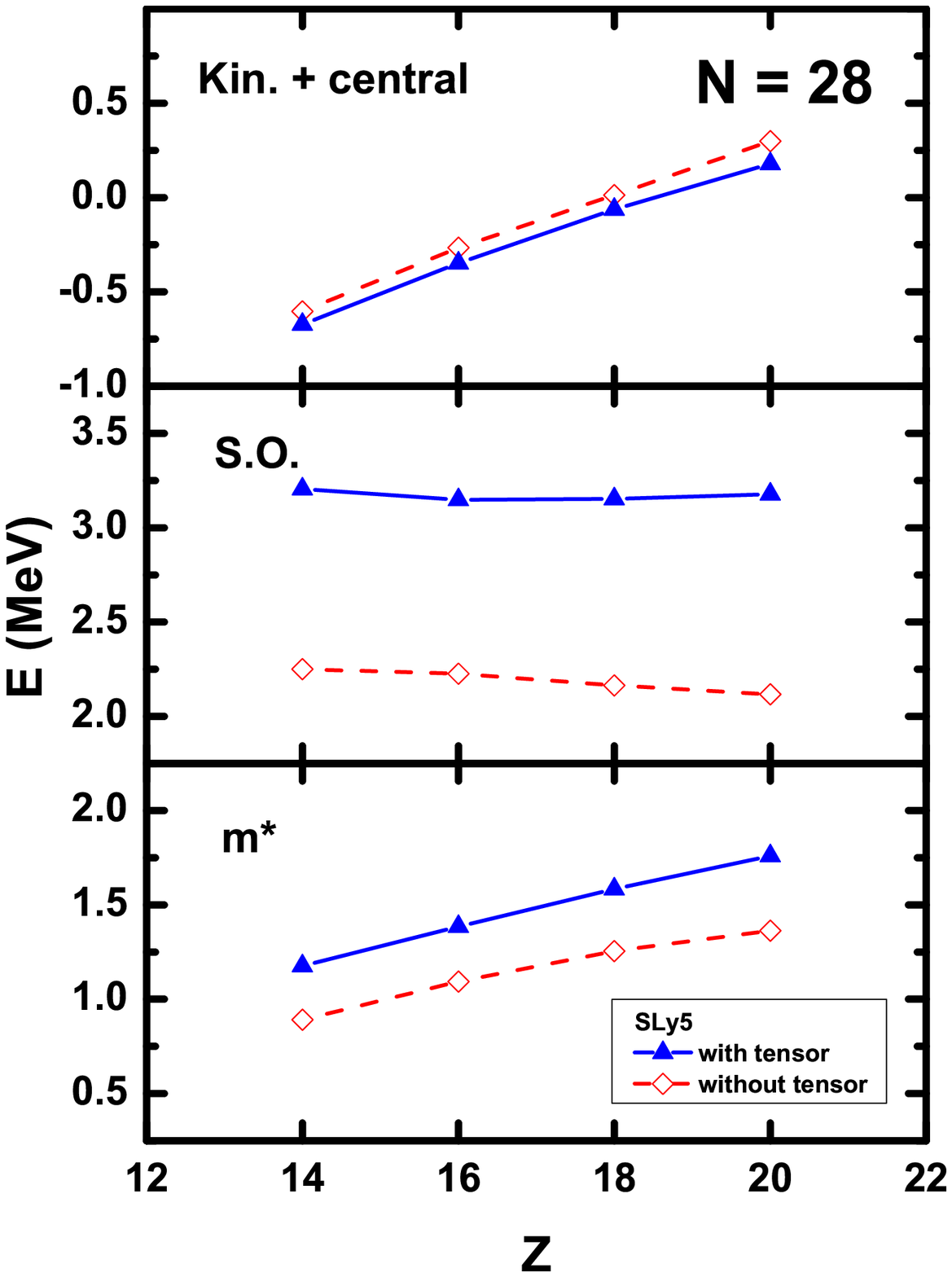}
\caption{The 3 contributions to $\Delta_{\rm{gap}}(Z)$ according to the
decomposition Eq.(\ref{MG4}). The results correspond to the Skyrme
interaction SLy5, with (solid lines) and without (dashed lines)
tensor component. } \label{Fig4}
\end{figure}

\begin{table}[h]
 \caption{\label{Tab:2} The $2p_{3/2}$-$1f_{7/2}$ gap for the isotonic chain $N=28$;
    The units are MeV.}
 \begin{ruledtabular}
 \begin{tabular}{ccccc}
 ($N$,$Z$) & (28,14) & (28,16) & (28,18) & (28,20)    \\
\hline
SLy5 & 2.55 & 3.09 & 3.49 & 3.75 \\
SLy5+T & 3.79 & 4.08 & 4.66 & 5.15 \\
PKO1 & 2.13 & 3.52 & 5.11 & 4.78 \\
PKO2 & 2.93 & 4.16 & 5.26 & 5.22 \\
 \end{tabular}
 \end{ruledtabular}
 \end{table}

The calculated values of the $N=28$ gap $\Delta_{\rm{gap}}(Z)$ are
displayed in Table~\ref{Tab:2}. It can be seen that all models predict a
substantial reduction of this gap when going from $Z=20$ ($^{48}$Ca)
to $Z=14$ ($^{42}$Si). The largest reduction is for the two models
PKO1(2.65 MeV) and PKO2 (2.29 MeV), whereas the Skyrme models SLy5 and SLy5 + tensor predict a reduction of 1.20 MeV and 1.36 MeV, respectively. These
values are consistent with the experimental estimate of 1 MeV. The
non-relativistic results also indicate that the tensor component
of the interaction does not play an essential role in this
reduction.
This may be supported by the fact that both PKO1 and PKO2 relativistic models
lead to very similar reduction.

One can have a closer look at the mechanism of this gap reduction by
performing for the non-relativistic models an analysis similar to
that of Subsec.III.A. In Fig.~\ref{Fig4} are shown the contributions to
$\Delta_{\rm{gap}}(Z)$ coming from the different terms of Eq.(\ref{MG4}).
Comparing the results of SLy5 and
SLy5 + tensor one can see that the tensor force has an overall
effect on the spin-orbit contribution to $\Delta_{\rm{gap}}(Z)$ but it
affects little the gap reduction from $Z=20$ to $Z=14$. The other two
terms practically determine the gap reduction. In the SLy5 + tensor case for example, the partial decrease from $Z=20$ to $Z=14$ due to the kinetic + central effect is 840 keV, while the non-locality effect is 564 keV. 

We stress again
that, in the present study the particle-vibration effects are still
missing and their effects on $\Delta_{\rm{gap}}(Z)$ remain to be
evaluated.

\section{Conclusion}

In this work we have studied the mechanisms which lead to a
reduction of the proton $1d_{3/2}$-$1d_{5/2}$ spin-orbit splitting
$\Delta_{\rm{so}}(N)$ with decreasing the mass number in the Si isotopic chain. We also investigate a quenching of the neutron
$2p_{3/2}$-$1f_{7/2}$ gap $\Delta_{\rm{gap}}(Z)$ as one goes from $^{48}$Ca towards $^{42}$Si. We have used the self-consistent mean field approach in
its non-relativistic version with the SLy5 parametrization, and
its relativistic covariant version with exchange (Fock) terms.

One of the goals of this study is to determine to which extent a
tensor component in the effective nucleon-nucleon interaction can
affect the values of $\Delta_{\rm{so}}(N)$ and $\Delta_{\rm{gap}}(Z)$. Our
main conclusions are: 1) the reduction of $\Delta_{\rm{so}}(N)$ when
going from $^{34}$Si to $^{42}$Si is mainly due to the presence of
a tensor component in the effective interaction. The magnitude of
the change in $\Delta_{\rm{so}}(N)$ is consistent with the empirical
observations. 2) On the other hand, the evolution of
$\Delta_{\rm{gap}}(Z)$ when $Z$ decreases from 20 to 14 does not depend
strongly on a tensor component in the interaction.
These conclusions can be reached by using either a non-relativistic Skyrme mean field approach, or a relativistic Hartree-Fock framework.

In the case of the Skyrme-Hartree-Fock model, we have performed a detailed analysis of the origin of the evolution of $\Delta_{\rm{gap}}(Z)$ and  $\Delta_{\rm{so}}(N)$. The reduction of the former can be traced back to changes occurring in the symmetry part of the central potential, to the
$m^*$-term of the Skyrme-HF model, and only
to a small extent to the spin-orbit potential.


It would be important to extend in future studies this type of
analysis to include effects such as particle-vibration coupling
which are beyond the present mean field models and which are known
to affect single-particle spectra.

\section{Acknowledgments}
This work is partly supported by the
the European Community project Asia-Europe Link in Nuclear Physics
and Astrophysics CN/Asia-Link 008 (94791),
by the Bulgarian Science Foundation (contract VUF06/05), CNRS(France) - NSFC(China) PICS program no. 3473, and CNRS(France) - MoE(Bulgaria) PICS program no. 4329.

\end{document}